\documentclass[12pt]{article}
\usepackage[dvips]{graphicx}
\usepackage{pdproc}

  \makeatletter 
  \def\@cite#1{[#1]} 
  \makeatother    
\begin{document}

\renewcommand{\thefootnote}{\alph{footnote}}

\title{
 Leptoquark Searches at the Tevatron
}

\author{RAIMUND STR\"OHMER (for the CDF and D\O\ collaborations)}

\address{Ludwig-Maximilians-Universit\"at Munich \\
 Am Coulombwall 1, 85748 Garching, Germany }

\abstract{
This article summarizes the status of the Tevatron Run II searches 
for leptoquarks in $p\bar{p}$ collisions at $\sqrt{s}=1.96 $ TeV.
The analyses discussed use datasets with integrated luminosities
ranging from 72 pb$^{-1}$ to 200 pb$^{-1}$.}

\normalsize\baselineskip=15pt

\section{Introduction}

The existence of leptoquarks (LQ), color triplet bosons with lepton
and quark quantum numbers, is predicted by several extensions of 
the Standard Model~\cite{LQ}. Limits on the proton lifetime, lepton flavor
violating decays, and flavor changing neutral currents lead to the
assumption that the leptoquarks exist in three generations each coupling
only to one lepton and one quark generation.
At hadron colliders leptoquark pairs can be produced via the strong
interaction. The production cross section of scalar leptoquarks
is therefore practically model independent.
Each leptoquark
can either decay into a quark and a charged lepton 
or  into a quark and a neutrino. 
The branching ratio $\beta$ of the leptoquark to decay into a charged
lepton and a quark depends  on the details of the model.
The limits will therefore be given either as a function of $\beta$
or for fixed values of  $\beta$.  
The combinations of the two leptoquark decays lead to distinctive
topologies which will be discussed in the following sections.
Since the region of interest for leptoquark masses is typically of
the order of 200 GeV or higher the decay products of the leptoquarks will 
have high energies.
If both leptoquarks decay into charged leptons one will observe two
leptons of the same flavor and two highly energetic jets. If one 
leptoquark decays into a charged lepton and the other into a
neutrino one will observe two jets, one lepton, and missing 
transverse energy.
Finally if both leptoquarks decay into neutrinos one will observe
two jets and missing transverse energy.

\section{First generation leptoquarks}

\subsection{Two jets and two electrons}
\begin{figure}[htb]
\begin{center}
\includegraphics*[width=5.2cm]{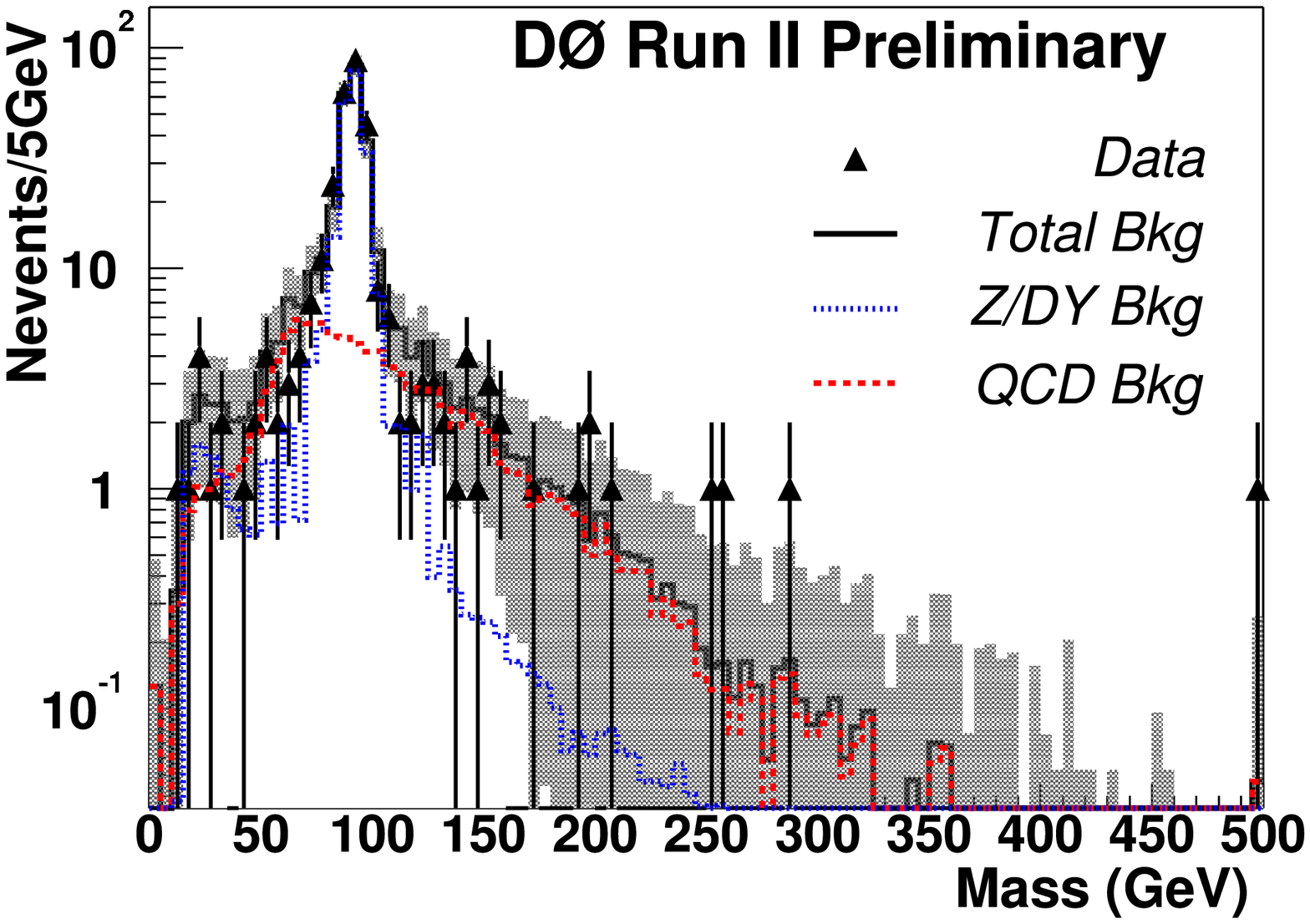}
\includegraphics*[width=5.2cm]{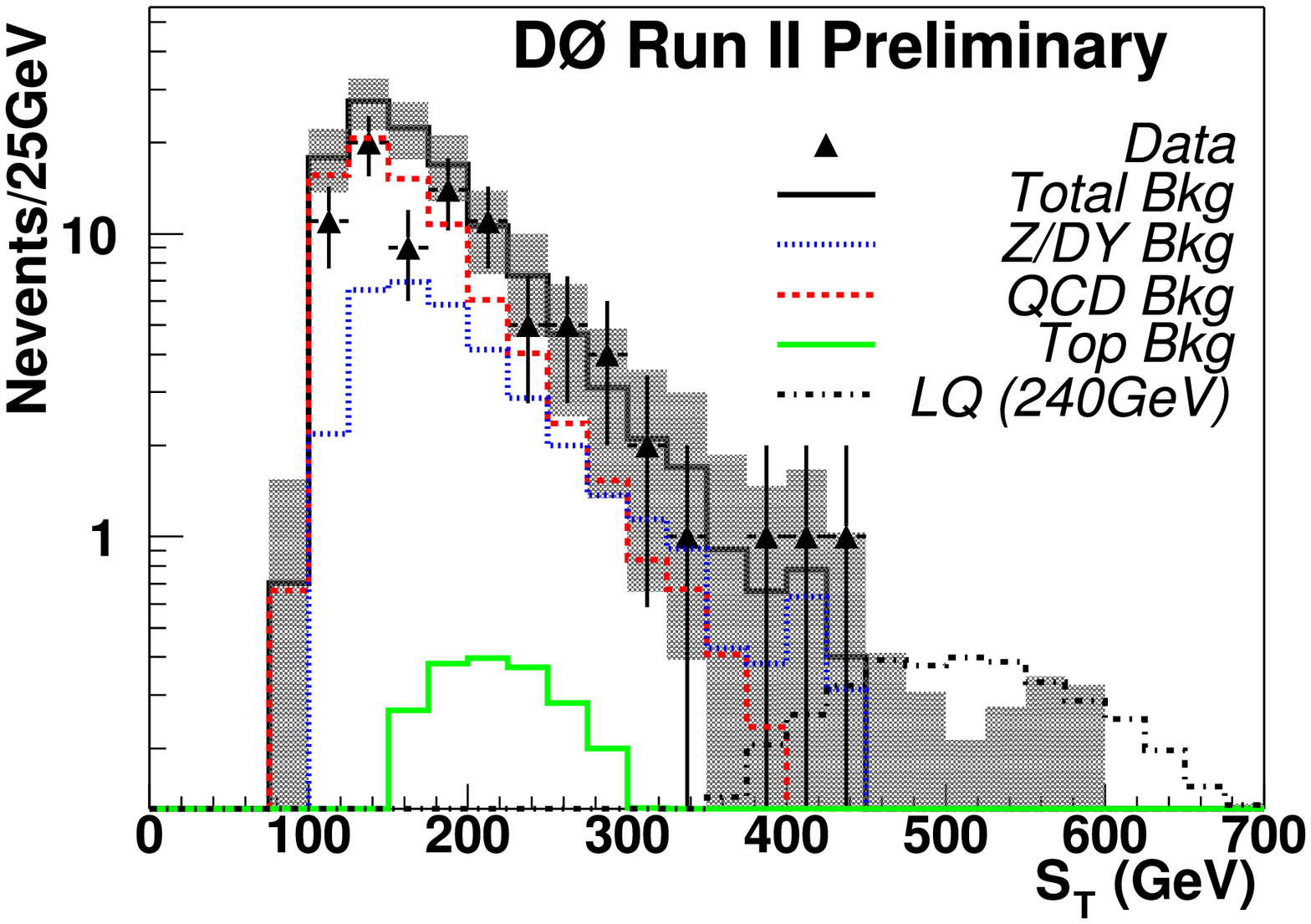}
\includegraphics*[width=5.2cm]{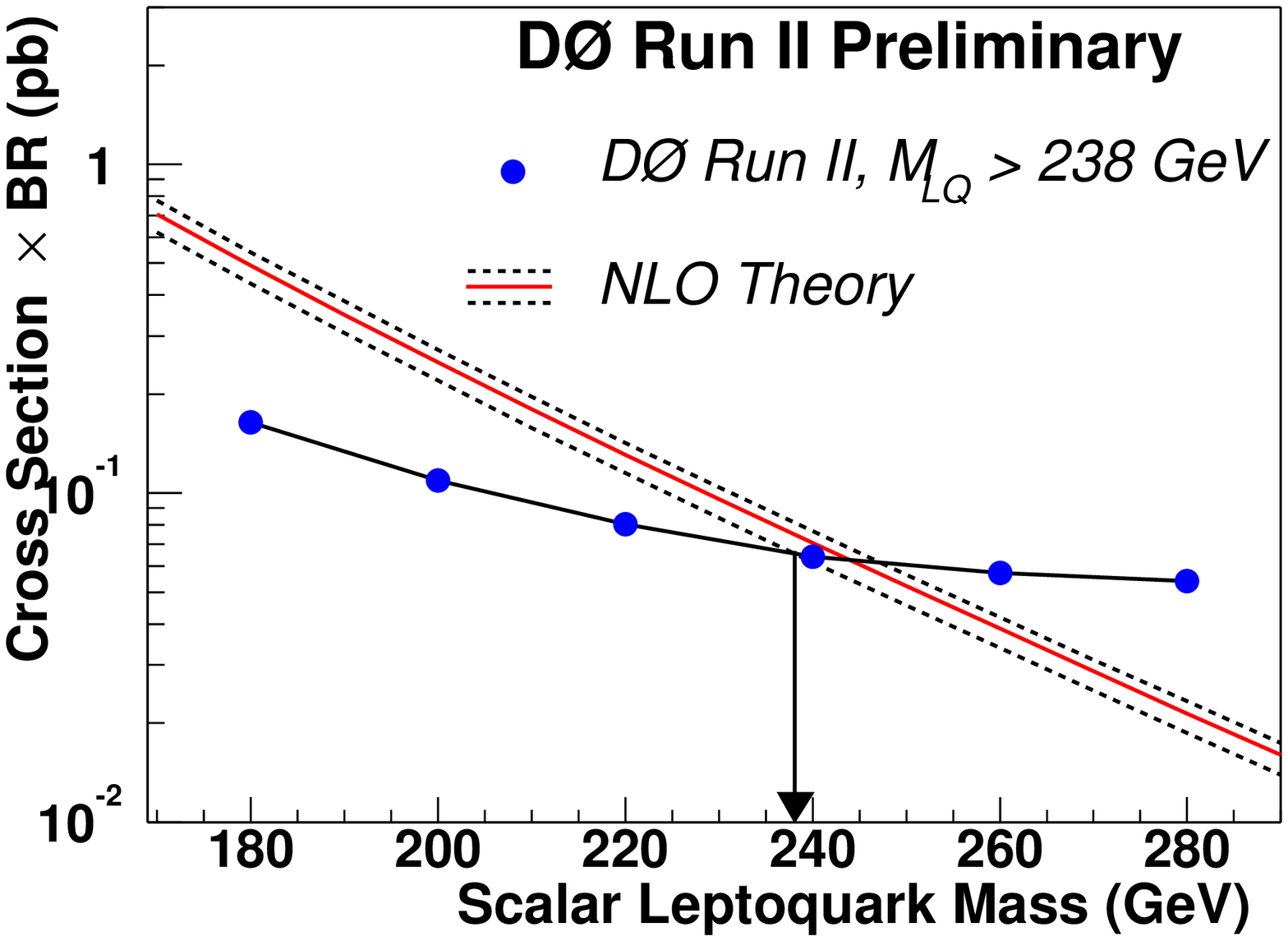}
\caption{%
Comparison of the electron pair mass and the $S_T$ distribution between 
the data and the expected background. The $S_T$ distribution is shown
after applying a Z-veto cut. The dot-dashed histogram in the $S_T$ plot
is the expectation for a 240 GeV leptoquark signal.
The shading represents the uncertainty on the total background.
Comparison of the 95\% confidence limit on the production cross section of
1$^{st}$ generation leptoquarks
with the theoretical NLO  cross section~\protect\cite{NLO}.
The dashed lines indicate the change in the NLO prediction if the
renormalization and factorization scales are changed 
from $M_{LQ}$ to $2M_{LQ}$ and to $M_{LQ}/2$. 
\label{fig:ee_dis}}
\end{center}
\end{figure}

If both leptoquarks decay into a quark and an electron one will
observe two high energetic electrons and two jets.
The basic selection for the preliminary D\O\ analysis are two
jets with transverse energy
$E_T>20$ GeV and two electrons with transverse energy $E_T>25$ GeV.
The main backgrounds are due to electron pairs from $Z$-decays
and Drell-Yan events with two additonal jets, and to
events where jets are misidentified as electrons. (In order to
maintain a very high signal efficiency the D\O\ analysis uses
loose electron selection criteria.)
The background can be suppressed by vetoing on the $Z$ mass region
in the electron pair mass and by exploiting the high energy of
the objects with a cut on the  scalar sum $S_T$ of the transverse energies
of the electrons and jets (see Fig.~\ref{fig:ee_dis}):
$S_T=E_T(e_1)+E_T(e_2)+E_T(j_1)+E_T(j_2)$. 

After all cuts the D\O\ analysis observes 0 events while $0.4\pm0.1$
events are expected from background. The typical efficiency is about 30\%.
Since no significant excess is observed a limit on the production
cross section times branching ratio can be calculated and
compared to the expected cross section
(see Fig.~\ref{fig:ee_dis}).
From this comparison D\O\ derives a preliminary limit on the mass 
of scalar leptoquarks of:
\begin{equation}
 M_{LQ}^{scalar}(1^{st}~gen)>238~GeV 
\mbox{\rm ( D\O\ preliminary, 175 pb$^{-1}$, $\beta=1$) }
\end{equation}
at 95 \% confidence level. The coresponding prelimary limit 
obtained by CDF with a similar analysis is:
\begin{equation}
 M_{LQ}^{scalar}(1^{st}~gen)>230~GeV 
\mbox{\rm ( CDF preliminary, 175 pb$^{-1}$, $\beta=1$). }
\end{equation}


\subsection{Two jets, one electron, and missing energy}
If one of the two leptoquarks decays into a neutrino and a quark and the
other into an electron and a quark one
will observe two jets, an electron, and missing energy due to the unobserved
neutrino. The main background in this case are
events with a W and two additional jets, and
events where a jet or photon is misidentified as an electron. D\O\ uses
a pre-selection of events with two jets with $E_T>25$ GeV, one electron 
with $E_T>35$ GeV, and
missing transverse energy  $E_T^{miss}>25$ GeV. Events containing W decays
are vetoed by requiring that the transverse mass of the electron and the
neutrino (estimated from the missing energy) be larger than 130 GeV. 
The high energy of the leptoquark decay products is again
exploited by a cut on the scalar sum of the four objects 
$S_T=E_T(e_1)+E_T^{miss}+E_T(j_1)+E_T(j_2)$.
The preliminary limits on the leptoquark mass for $\beta=0.5$ are:
\begin{equation}
 M_{LQ}^{scalar}(1^{st}~gen)>194~GeV 
\mbox{\rm ( D\O\ preliminary, 175 pb$^{-1}$, $\beta=0.5$), }
\end{equation}
\begin{equation}
 M_{LQ}^{scalar}(1^{st}~gen)>166~GeV 
\mbox{\rm ( CDF preliminary, 72 pb$^{-1}$, $\beta=0.5$). }
\end{equation}

\subsection{Two jets and missing energy}
\begin{figure}[htb]
\begin{center}
\includegraphics*[width=6.5cm]{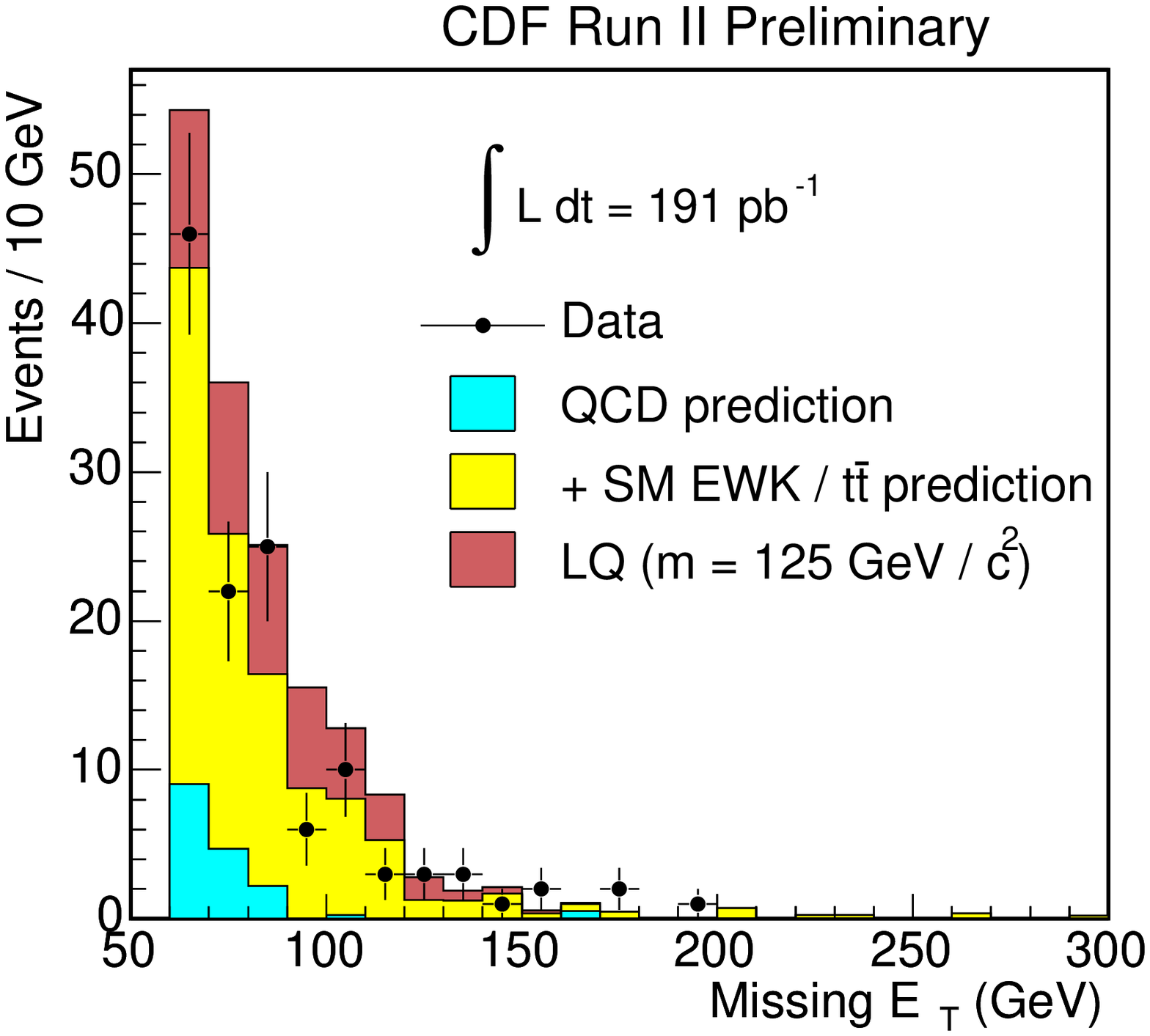}
\includegraphics*[width=5.4cm]{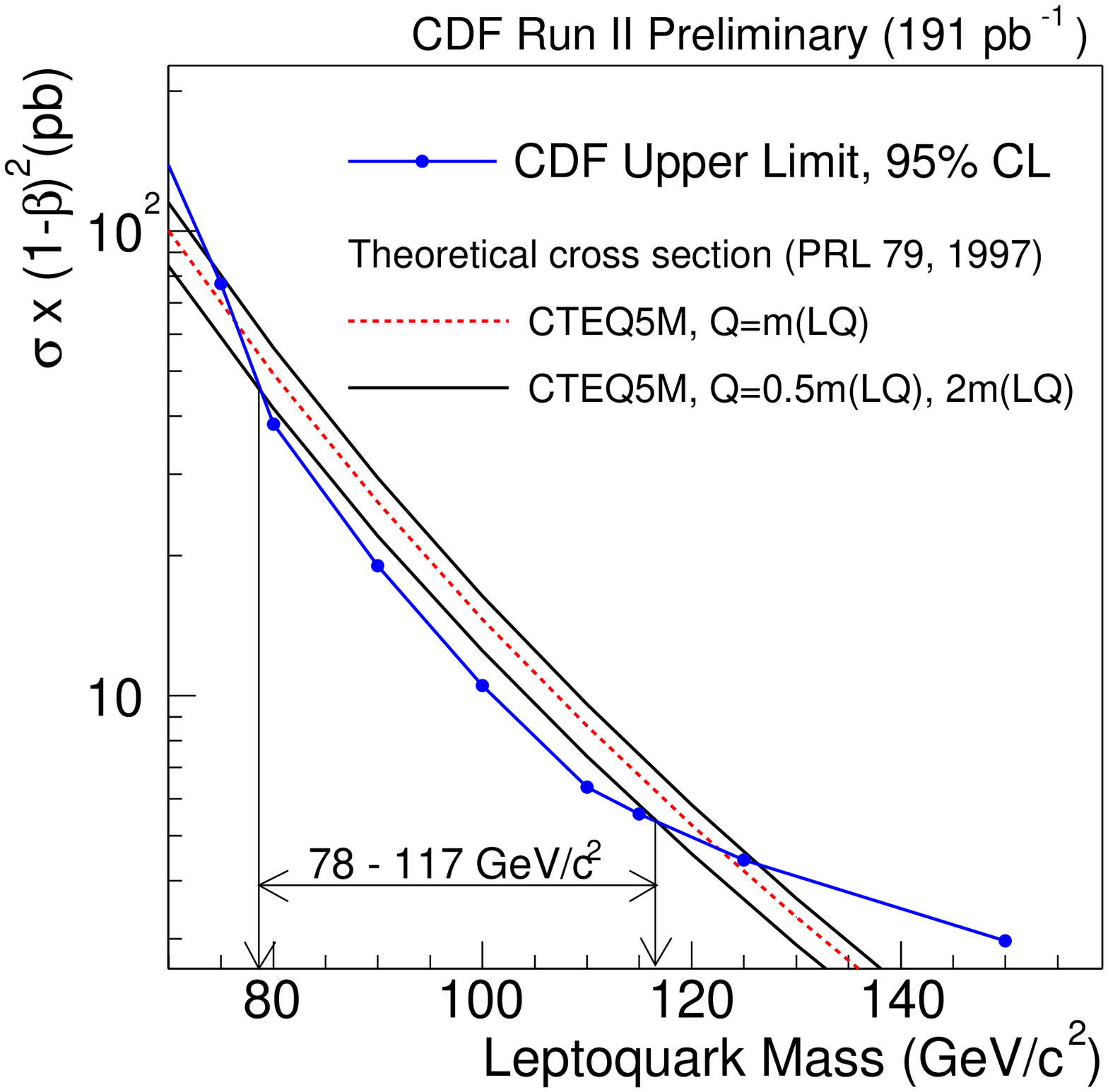}
\caption{%
Missing transverse energy in the leptoquark signal region.
Comparison of the 95\% confidence limit on the production cross section of
leptoquarks with the  theoretical NLO  cross section.
\label{fig:top}}
\end{center}
\end{figure}

If both leptoquarks decay into  a neutrino and a quark only two
jets and missing energy are observed. The CDF analysis
requires two jets with $E_T>20$ GeV and missing transverse energy
above 60 GeV. After all cuts 124 events are observed with an expected
background of $118\pm14$ events. For $\beta=0$ this excludes scalar 
leptoquarks in the mass range between 78 GeV and 117 GeV at 95\%
confidence level (see Fig.~\ref{fig:top}).
This limit is independent of the leptoquark generation.

\subsection{Combination}
\begin{figure}[htb]
\begin{center}
\includegraphics*[width=8.2cm]{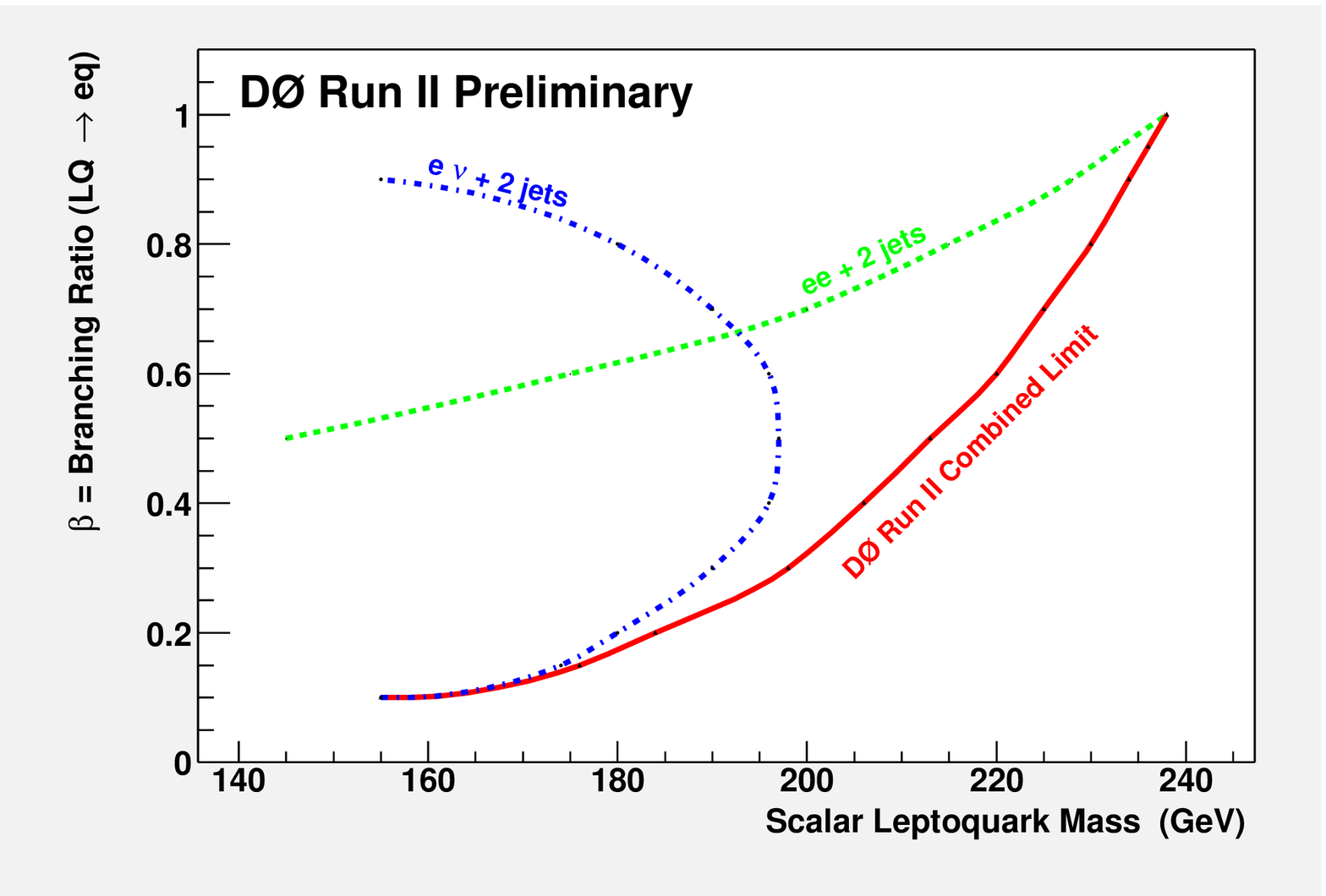}
\includegraphics*[width=6.0cm]{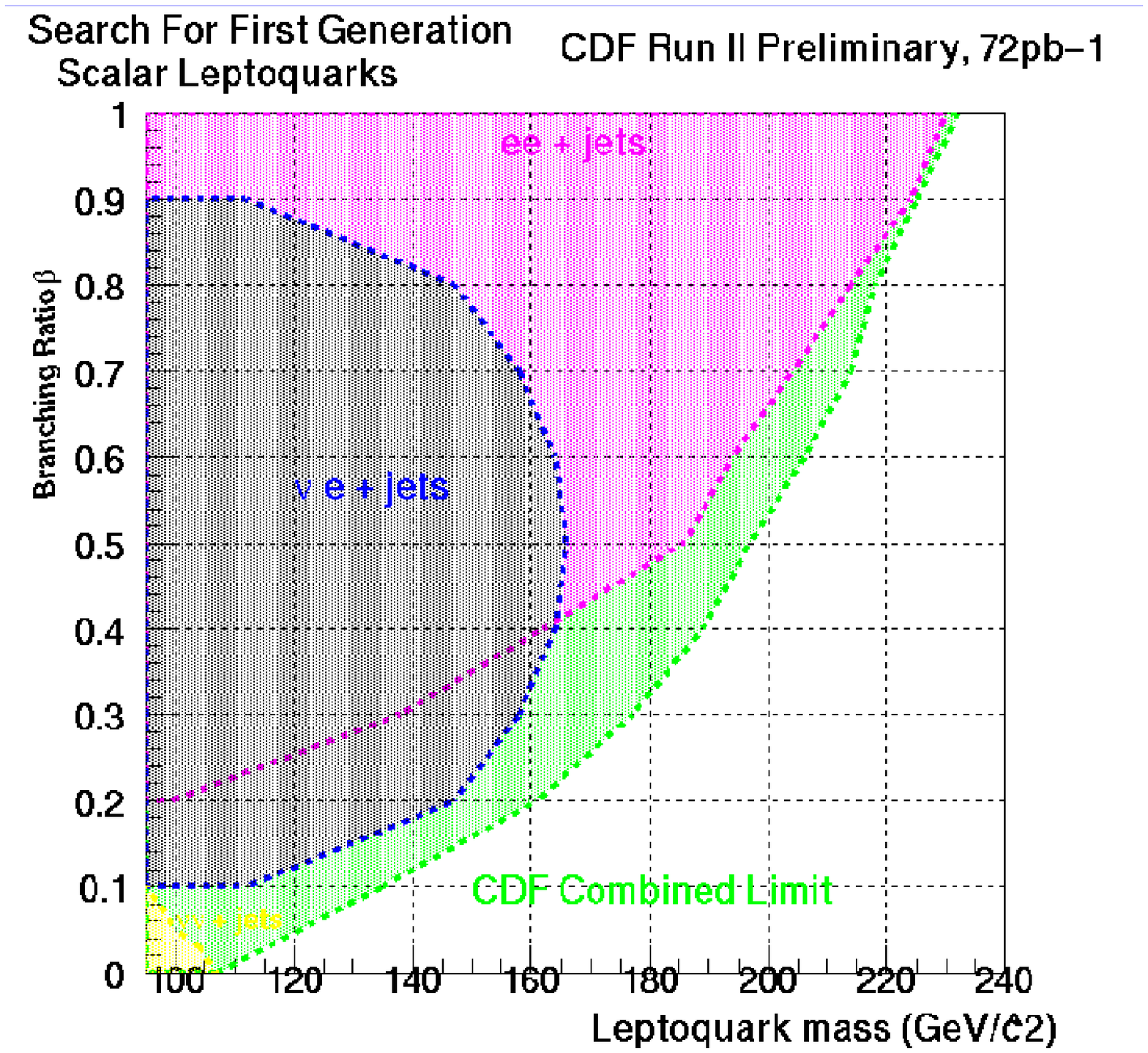}
\caption{%
95\% CL lower limits on the mass of a first generation scalar leptoquark
as a function of $\beta$.
\label{fig:e_lim}}
\end{center}
\end{figure}

Fig~\ref{fig:e_lim} shows the 95\% confidence level lower limit on the
mass of a first generation scalar leptoquark  as a function of 
decay probability $\beta$ of the leptoquark into an electron and a quark.
In particular for  $\beta=0.5$ D\O\ finds:
\begin{equation}
 M_{LQ}^{scalar}(1^{st}~gen)>213~GeV 
\mbox{\rm ( D\O\ preliminary, 175 pb$^{-1}$, $\beta=0.5$). }
\end{equation}

\section{Second generation leptoquarks}

For second generation leptoquarks both collaborations have studied
the case where both leptons decay into a muon and a quark.
The main backgrounds are $Z$ decays and Drell-Yan events with two
additional jets. The CDF analysis uses a pre-selection of two muons 
with $p_T>25$ GeV and two jets with $E_T^{j1}>30$ GeV and $E_T^{j2}>15$ GeV.
Events with a muon pair mass of $M_{\mu\mu}<15$ GeV or 
$75~\mbox{\rm GeV} < M_{\mu\mu} < 105$ GeV are vetoed. The background is 
finally reduced by a two dimensional cut on $E_T(j_1)+E_T(j_2)$ and
$p_T(\mu_1)+p_T(\mu_2)$.

The preliminary results are:
\begin{equation}
 M_{LQ}^{scalar}(2^{nd}~gen)>241~GeV 
\mbox{\rm ( CDF preliminary, 198 pb$^{-1}$, $\beta=1.$) }
\end{equation}
\begin{equation}
 M_{LQ}^{scalar}(2^{nd}~gen)>186~GeV 
\mbox{\rm ( D\O\ preliminary, 104 pb$^{-1}$, $\beta=1.$) }
\end{equation}

\section{Conclusions}
Hadron colliders where leptoquarks can be produced  in pairs via the
strong interaction are good places to search for leptoquarks.
The large energies of the decay products of the leptoquarks provide
clear signatures which can be distinguished from the background.
Both D\O\ and CDF have searched for first and second generation leptoquarks,
But no evidence for leptoquarks has been found. Therefore preliminary limits
on the leptoquark mass are presented. These limits
surpass the Run I limits.

\section{Acknowledgements}
I would like to thank my colleagues from the D\O\ and CDF 
collaborations for providing these excellent results.

\bibliographystyle{plain}

\begin{thebibliography}{99}
\bibitem{LQ} 
D. Acosta and S. Blessing, Annu. Rev. Nucl. part. Sci. 49, 389(1999), 
and references therein.
%
\bibitem{NLO}
M. Kramer {\it et. al.}, Pair Production of Scalar Leptoquarks at the Fermilab
Tevatron, Phys. Ref. Lett. 79, 341(1997). 
%
\end{thebibliography}

\end{document}